\documentclass[pre,twocolumn,epsfig,rotate,showpacs]{revtex4}
\usepackage{epsfig}
\usepackage{graphicx}
\usepackage{amsmath}

\newcommand{\beq}{\begin{eqnarray}}
\newcommand{\eeq}{\end{eqnarray}}
\begin{document}

\title{Generating a Fractal Butterfly Floquet Spectrum in a Class of Driven SU(2) Systems}
\author{Jiao Wang$^{1,2}$ and Jiangbin Gong$^{3,4}$}
\email{phygj@nus.edu.sg} \affiliation{$^{1}$Temasek Laboratories,
National University of Singapore, 117542, Singapore
\\$^{2}$Department of Physics, Institute of Theoretical Physics and Astrophysics,
Xiamen University, Xiamen 361005, China
\\$^{3}$Department of Physics and Center of Computational Science and Engineering,
National University of Singapore, 117542, Singapore
\\$^{4}$NUS Graduate School for Integrative Sciences and Engineering, Singapore
 117597, Singapore}
\date{\today}

\begin{abstract}

A scheme for generating a fractal butterfly Floquet spectrum, first
proposed by Wang and Gong [Phys. Rev. A {\bf 77}, 031405(R) (2008)],
is extended to driven SU(2) systems such as a driven two-mode
Bose-Einstein condensate.  A new class of driven systems without a
link with the Harper model context is shown to have an intriguing
butterfly Floquet spectrum.  The found butterfly spectrum shows
remarkable deviations from the known Hosftadter's butterfly. In
addition, the level crossings between Floquet states of the same
parity and between Floquet states of different parities are studied
and highlighted. The results are relevant to studies of fractal
statistics, quantum chaos, coherent destruction of tunneling, as
well as the validity of mean-field descriptions of Bose-Einstein
condensates.

\end{abstract}

 \maketitle

\section{Introduction}

Hofstadter's butterfly spectrum of the Harper model \cite{hof},
first discovered in two-dimensional electron systems subject to a
square lattice potential and a perpendicular magnetic field, has
attracted tremendous mathematical, theoretical and experimental
interests. For an arbitrary irrational value of one system
parameter, the spectrum of the Harper model is a fractal, which
has been strictly proved after decades of research on the ``Ten
Martini problem" \cite{simon}. As one important property of
Hofstadter's butterfly spectrum, the number of its sub-bands
depends on the arithmetic property of the flux of the magnetic
field. As the magnetic flux changes smoothly and thus varies
between irrational or rational numbers, the gap between the
sub-bands shows fractal properties and will close itself
infinitely times \cite{goldman}. This implies that Hofstadter's
butterfly spectrum contains infinite quantum phase transition
points \cite{phasebook}.

Early quantum chaos studies established that the Floquet
(quasi-energy) spectrum of periodically driven systems may display a
fractal butterfly pattern as well \cite{kickedharper,jiao08,jiao09}.
However, the nature of the fractal Floquet spectrum is still poorly
understood for three reasons. First, because the eigen-phase of
Floquet states is restricted to a range of $2\pi$, understanding a
Floquet spectrum associated with an infinite-dimensional Hilbert
space is subtle and challenging \cite{kohn}. Second, a rigorous
mathematical proof about the fractal nature of a butterfly Floquet
spectrum is still lacking. Third, previous findings regarding to
fractal Floquet spectrum were largely limited to the so-called
kicked-Harper model (a driven version of the Harper model)
\cite{casatietal,dana1,dana2} and its variant
\cite{jiao08,jiao09,lawton,dana2}.

Because of great interests in studies of quantum control, especially
in studies of dressed matter waves
\cite{holthaus,arimondo,Korsch,Qi,weiss}, there are now promising
possibilities for the engineering and simulation of driven ultracold
systems with a prescribed Floquet spectrum. Reference \cite{jiao08}
represents a recent attempt in this direction. In particular, in
Ref. \cite{jiao08} we showed that by designing two $\delta$-kicking
sequences the Floquet spectrum of a double-kicked-rotor system can
be made to be a Hofstadter's butterfly, and the spectrum is
identical with that of a kicked-Harper model when a certain
parameter takes an arbitrary irrational
value \cite{lawton}. 

In this paper, we reveal yet another class of butterfly Floquet
spectrum using driven SU(2) systems, which are realizable by, for
example,  a driven two-mode Bose-Einstein condensate (BEC).  As seen
below, the basic strategy is essentially an extension of our
previous work \cite{jiao08}, thus suggesting the possibility of
synthesizing butterfly spectrum in many other systems that go well
beyond the context of two-dimensional electron systems or the
Harper-model context.

Our findings about the butterfly spectrum of  driven SU(2) systems
are both motivating and fascinating.  First of all, as explained
below, now all the three popular paradigms of quantum chaos, i.e.
the kicked-rotor model, the kicked-Harper model, and the kicked-top
model, are linked together, insofar as any one of them can be used
to generate quantum critical systems with fractal statistics.
Second, the butterfly spectrum obtained in driven SU(2) systems is
significantly different from Hofstadter's butterfly, with remarkable
aspects. For example, we show that with one certain system parameter
fixed the overall butterfly pattern is insensitive to the number of
bosons (denoted $N$) in the BEC, but some detailed features depend
on whether $N$ is odd or even. This may serve as a useful guide for
seeking dramatic coherence effects in a BEC. Another interesting
aspect is that the butterfly spectrum contains many level crossings
between states of different parities and thus many points of
coherent destruction of tunneling (CDT) \cite{Hanggi}, with the
total number of CDT points found to scale as $\sim N^{3.0}$. As an
analog of  quantum phase transitions in driven systems, the found
butterfly pattern also contains many level crossings between
same-parity eigenstates. Due to these distinctive properties, the
butterfly spectrum reported here represents a new class of fractal
Floquet spectrum and may become a test bed for a number of research
topics. To emphasize the remarkable differences between the
butterfly spectrum found here and that associated with Harper's
model,  we refer to the newly found spectrum as ``butterfly
spectrum'' instead of ``Hofstadter's butterfly spectrum''.

The main results of this study has been briefly reported in Ref.
\cite{WGprl} and this paper represents a full-length description
of our findings. In Sec. II we will introduce the model SU(2)
system and explain the main idea behind our study. In Sec. III we
study the peculiar multifractal spectral properties of the
butterfly spectrum and the associated level crossings.  The
relevance of the underlying classical limit is also discussed in
detail.  To motivate possible experiments, we discuss some related
issues in Sec. IV.  We conclude this study in Sec. V. Appendices
A, B, and C present some further details that may be of interest
to some readers.

\section{Driven SU(2) Model}

Our driven SU(2) model was motivated by a driven two-mode BEC
system, proposed earlier \cite{Milburn,Korsch,liujie} to realize the
well-known kicked-top model \cite{Haakebook} in the quantum chaos
literature. In a very general form, a driven two-mode Bose-Hubbard
model can be written as
\begin{eqnarray}
H=f(t)\hbar (a_{1}^{\dagger}a_2+ a_{2}^{\dagger}a_1)+ g(t)\hbar
(a_{1}^{\dagger}a_1-a_{2}^{\dagger}a_2)^2, \label{Hami}
\end{eqnarray}
where $a_{i}$ and $a_{i}^{\dagger}$ are the bosonic annihilation
and creation operators for the $i$th mode, $f(t)$ describes the
time-dependent tunnelling rate between the two modes, and the
$g(t)$ term describes the self-interaction between same-site
bosons, whose time dependence can be achieved by Feshbach
resonance induced by an additional magnetic field. Note that the
total number of bosons $N=a_{1}^{\dagger}a_1+a_2^{\dagger}a_2$ is
a conserved quantity. For a fixed $N$, the dimension of the
Hilbert space is $N+1$. Using the Schwinger representation of
angular momentum operators, namely, $J_{x}=
(a_{1}^{\dagger}a_2+a_{2}^{\dagger}a_1)/2$,
$J_y=(a_{2}^{\dagger}a_1-a_{1}^{\dagger}a_2)/(2i)$, and
$J_z=(a_{1}^{\dagger}a_1-a_{2}^{\dagger}a_2)/2$, Eq. (\ref{Hami})
reduces to
\begin{eqnarray}
H= 2f(t)\hbar J_x+ 4g(t)\hbar J_z^2. \label{Hami2}
\end{eqnarray}
This above Hamiltonian makes it clear that its dynamics is solely
determined by the SU(2) generators $J_x$, $J_y$ and $J_z$. The
total angular-momentum quantum number $J$ is given by $J=N/2$. The
Hilbert space can be expanded by the eigenstates of $J_z$, denoted
$|m\rangle$, with $J_z|m\rangle=m|m\rangle$.  The population
difference between the two modes is given by the expectation value
of $2J_{z}$.  It is also important to note that if we exchange the
indices of the two modes, then $J_x$ is invariant, $J_z\rightarrow
-J_z$, and as a result the Hamiltonian in Eq. (\ref{Hami2}) is
unchanged. This reflects a parity symmetry of our model, which
will be exploited below.

Consider then two specific forms of $f(t)$ and $g(t)$. In the first
case $f(t)=\alpha/(2\tau)$, $g(t)= g_0
\sum_{n}[\delta(t-2n\tau-\tau)-\delta(t-2n\tau)]$. The Floquet
operator, i.e., the unitary evolution operator $F$ from
$2n\tau+0^{+}$ to $(2n+2)\tau+0^{+}$, is then given by
\begin{eqnarray}
F=e^{i\eta J_z^2/(2J)}e^{-i \alpha J_x} e^{-i\eta J_z^2/(2J)}
e^{-i\alpha J_x}, \label{Feq}
\end{eqnarray}
where $\eta=4 g_0 N$.  Interestingly, the first two or the last two
factors in Eq. (\ref{Feq}) constitute the Floquet operator for a
standard kicked-top model \cite{Haakebook}. As such our driven
system here can be regarded as a ``double-kicked-top model".
Alternatively, if we set $g(t)=g_0/\xi$,
$f(t)=\frac{\alpha}{2}\sum_{n}[\delta(t-n\tau)+\delta(t-n\tau-\xi)]$,
where $\xi$ is the time delay between the two delta kicking
sequences, then the associated propagator $F'$ from $n\tau-0^{+}$ to
$(n+1)\tau-0^{+}$ is given by
\begin{eqnarray}
F'=e^{-i(4g_0\tau/\xi)J_z^2}e^{i\eta J_z^2/(2J)}e^{-i\alpha J_x}
e^{-i\eta J_z^2/(2J)} e^{-i\alpha J_x}. \label{2ndF}
\end{eqnarray}
Under the special condition $4g_0\tau/\xi=2k\pi$ ($8k\pi$) for
integer $J$ (half integer $J$), where $k$ is an integer , the factor
$e^{-i(4g_0\tau/\xi) J_z^2}$ is unity in the $(2J+1)$-dimensional
Hilbert space and hence $F'$ becomes identical with $F$.  Based on
this, one now has two different scenarios for realizing $F$, the key
operator to be analyzed below.

\begin{figure}
\vspace{-.2cm}\hspace{0.cm}\epsfig{file=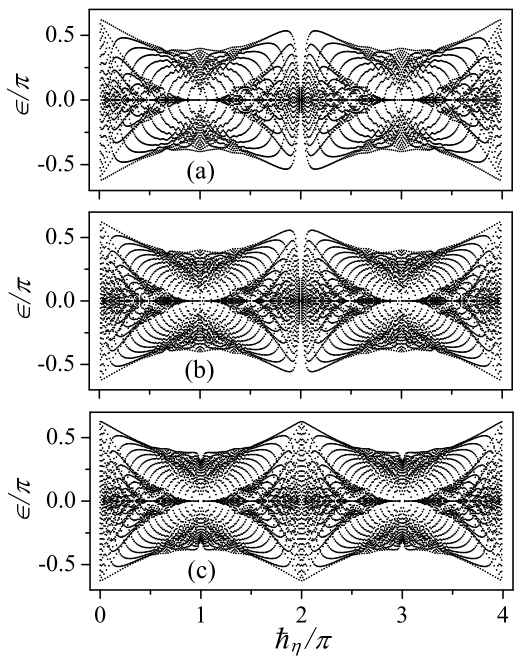,width=7.5cm}
\vspace{-1.7cm}\caption{The eigen-phase spectrum (denoted
$\epsilon$) of the Floquet operator $F$ in Eq. (3). $J=20$ in (a),
30 in (b), and 30.5 in (c). $\alpha/\hbar_{\text{eff}}=1$ in all
panels. Beautiful butterfly patterns are clearly recognized.
Analogous result for $J=100$ can be found in Ref.
\cite{WGprl}.}\label{sptm1}
\end{figure}

To explain our motivation of considering the $F$ operator, let us
consider the $|m\rangle$ representation. In that representation the
third factor $e^{-i\eta J_z^2/(2J)}$ of $F$ equals $e^{-i\eta
m^2/(2J)}$, which is a pseudo-random number for irrational $\eta/J$.
Interestingly, the first factor of $F$ however effectively induces a
time-reversal of the third factor and thus partially cancels this
pseudo-random phase. Indeed, using the SU(2) algebra \cite{fox}, the
product of the first three factors of $F$ in Eq. (\ref{Feq}) is
given by
\begin{eqnarray}
& & e^{i\eta J_z^2/(2J)}e^{-i \alpha J_x} e^{-i\eta
J_z^2/(2J)}\nonumber
\\
&=& e^{-i\alpha\left \{(J_x/2+iJ_y/2)e^{i[\eta (2J_z+1)/(2J)]} +
\text{c.c.}\right\}}.\label{rdmphase}
\end{eqnarray}
This shows that the $\eta$-dependent term entering into $F$ becomes
$e^{i[\eta (2J_z+1)/(2J)]}$, which is always a quasi-periodic number
$e^{i[\eta (2m+1)/(2J)]}$ in the $|m\rangle$ representation.
According to our early work \cite{gong07}, such a partial
cancelation of quasi-random dynamical phases implies intriguing
spectral properties.

For later discussion we also study the classical limit of $F$. To
that end we consider scaled variables $x=J_x/J$, $y=J_y/J$, and
$z=J_z/J$. Evidently, the three operators $x$, $y$, $z$ also satisfy
the angular-momentum algebra, but with an effective Planck constant
$\hbar_{\text{eff}}\equiv 1/J$. Taking the $\hbar_{\text{eff}}
\rightarrow 0$ limit with fixed $\eta$ and $\alpha$, the classical
dynamics associated with $F$ can be obtained, with variables $x$,
$y$, and $z$ restricted on a unit sphere. Because $\eta=4g_0 N$,
this classical limit with fixed $\eta$ requires $N\rightarrow
+\infty$ and $g_0\rightarrow 0$. This condition is apparently
equivalent to that in a standard mean-field limit of the driven BEC.

In addition to the system defined by $F$, we also consider one of
its interesting variants:
\begin{eqnarray}
F_{xy}=e^{i\eta J_z^2/(2J)}e^{-i \alpha J_x} e^{-i\eta J_z^2/(2J)}
e^{-i\alpha J_y}, \label{Fxy}
\end{eqnarray}
which is different from $F$ in the last factor, i.e., $e^{-i\alpha
J_x}$ in $F$ is replaced by $e^{-i\alpha J_y}$. As seen below, such
a variant may induce considerable changes in the spectral
properties. 

\begin{figure}
\vspace{-.2cm}\hspace{0.cm}\epsfig{file=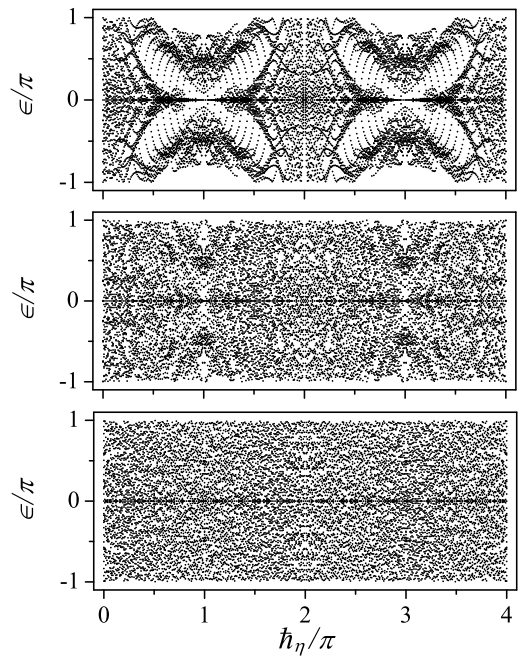,width=7.5cm}
\vspace{-1.7cm}\caption{The dissolving of the butterfly pattern of
the eigen-phase spectrum of the Floquet operator $F$ in Eq. (3). In
all the panels $J=30$. From top to bottom:
$\alpha/\hbar_{\text{eff}}=3$, $6$, and $10$, respectively. This
process resembles that observed in the kicked Harper model
\cite{kickedharper}.}\label{sptm2}
\end{figure}

\section{Detailed Aspects of the Butterfly Spectrum}

\subsection{Multifractal properties}

In the $|m\rangle$ representation, the matrix elements of the
operator $F$ can be evaluated in a straightforward manner.
Diagonalizing $F$ numerically then yields its spectrum.   Figure
\ref{sptm1} shows the typical eigen-phase spectrum of $F$ vs
$\hbar_{\eta}\equiv \eta \hbar_{\text{eff}}=\eta/J=8g_0$, for
$J=20, 30, 30.5$ and $\alpha/\hbar_{\text{eff}}=\alpha J=1$.
Because the spectrum of $F$ is invariant if
$\hbar_{\eta}\rightarrow \hbar_{\eta}+4\pi$ (see the proof in
Appendix A), we set $\hbar_{\eta}\in[0, 4\pi)$. Though in Fig.
\ref{sptm1} the involved Hilbert space is rather small,
spectacular butterfly patterns are already obtained. Their
reflection symmetry with respect to $\hbar_\eta=2\pi$ is also
clearly seen, a fact proved in Appendix B. The found butterfly
patterns in Fig. 1 resemble the famous Hofstadter's butterfly, but
also present remarkable differences in several aspects. First, if
we take a vertical cut of the butterfly patterns in Fig.
\ref{sptm1}, the spectrum is not found to present any large gaps.
Second, the butterfly patterns shown in each panel of Fig.
\ref{sptm1} possess a double-butterfly structure, with each
butterfly covering a $2\pi$ range of $\hbar_{\eta}$. This
double-butterfly structure is somewhat analogous to the spectrum
of a Harper-like effective Hamiltonian considered in Ref.
\cite{dana2}. More interestingly, though Fig. \ref{sptm1}(b)-(c)
has more levels than Fig. \ref{sptm1}(a), the overall outline of
the double-butterfly structure is seen to be insensitive to $J$
for fixed $\alpha/\hbar_{\text{eff}}=\alpha J$. Indeed, in Fig.
1(c) of Ref. \cite{WGprl} we also presented the spectrum for a
much larger $J$ value, i.e., $J=100$, and again similar outline of
the butterfly spectrum is obtained. Qualitatively this is because
when $\alpha/\hbar_{\text{eff}}=\alpha J$ is fixed, the phase
range of the second and fourth factors of $F$ is also fixed. By
contrast, for a fixed value of $J$ but for other not too large
values of $\alpha$, the qualitative features of the butterfly
spectrum remain, but at different scales. For very large values of
$\alpha$ (e.g., $\alpha/\hbar_{\text{eff}}>10$), the butterfly
pattern for a fixed value of $J$ will gradually dissolve, as seen
in Fig. \ref{sptm2}. This dissolving process of a butterfly
spectrum is similar to that seen in the kicked-Harper model
\cite{kickedharper}.

\begin{figure}
\vspace{-0.1cm}\epsfig{file=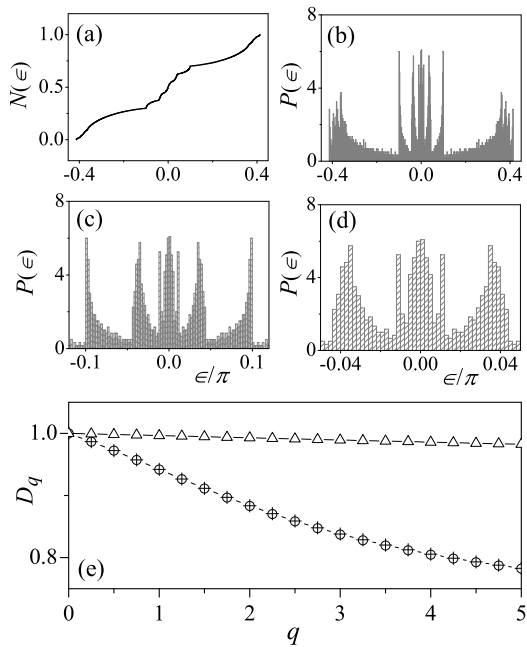,width=7.6cm}
\vspace{-2.1cm}\caption{Cumulative Floquet state density (a) and the
Floquet state density distribution (b)-(d) at different scales, for
$\hbar_{\eta}=(\sqrt{5}-1)\pi/2$, $\alpha/\hbar_{\text{eff}}=1$, and
$J=2999$. Panel (e) shows the generalized fractal dimension $D_q$.
Crosses and circles are for odd-parity and even-parity states.
Triangles represent the result for a standard kicked-top
model.}\label{Dq}
\end{figure}

Some detailed features of the spectrum are also noteworthy.  For
example, it is observed that the spectrum collapses to one point for
$\hbar_{\eta}=2\pi$, if and only if $J$ is an integer. This can be
explained as follows. If $J$ is an integer and if
$\hbar_{\eta}=2\pi$, then in the $|m\rangle$ representation,
\begin{eqnarray}
e^{-i\eta J_z^2/(2J)}&=&e^{-i\pi m^2} \nonumber \\
&=&e^{-i\pi m} \nonumber \\
&=& e^{-i\pi J_z}. \end{eqnarray}
 So in this case $e^{-i\eta
J_z^2/(2J)}$ is equivalent to a rotation of $\pi$ around the $z$
axis, and hence the first three factors of $F$ exactly cancel its
last factor. This cancellation will not occur if $J$ is a half
integer, i.e., if $N$ is odd.  Later we will return to this
intriguing difference between odd-$N$ and even-$N$ cases.

\begin{figure}
\vspace{-0.2cm}\epsfig{file=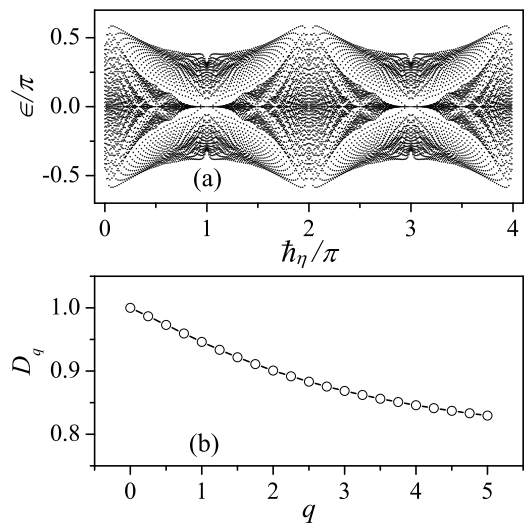,width=7.6cm}
\vspace{-3.3cm}\caption{(a) Eigen-phase spectrum of the Floquet
operator $F_{xy}$ (Eq. (\ref{Fxy})) for $J=30$. (b) The
corresponding fractal dimension $D_q$ computed for $J=2999$ and
$\hbar_{\eta}=(\sqrt{5}-1)\pi/2$. $\alpha/\hbar_{\text{eff}}=1$ in
both panels.}\label{Dqxy}
\end{figure}

We have also examined the statistical behavior of the found
butterfly spectrum. To have good statistics we consider a much
larger $J=2999$. Figure \ref{Dq}(a) presents the cumulative level
density $N(\epsilon)$ for a representative value of
$\hbar_{\eta}$. It is seen that $N(\epsilon)$ is highly irregular,
but does not show any clear flat steps. This is consistent with
our early observation that no large gap exists in the spectrum.
Figure \ref{Dq}(b)-(d) shows the associated level distribution
$P(\epsilon)$ at three different scales. Evidently, $P(\epsilon)$
has a fascinating self-similar property. This motivates us to
quantitatively characterize the spectrum via the generalized
fractal dimension $D_{q}$, with the results shown in Fig.
\ref{Dq}(e). As expected from the $N(\epsilon)$ result in Fig.
\ref{Dq}(a), $D_0=1$. However $D_q$ for $q\ne 0$ clearly shows
that the spectrum has multifractal properties. For the sake of
comparison, Fig. \ref{Dq}(e) also shows the $D_q$ result for a
standard kicked-top model with the same values of $\eta$ and
$\alpha$ (i.e., considering an operator comprising only the first
two factors of $F$). The $D_q$ behavior in the kicked-top case is
as trivial as that of a random sequence: it remains close to unity
and slightly decreases with increasing $q$ due to finite-size
effects. Based on these results, we conjecture and invite a formal
mathematical proof that the butterfly patterns found here contain
true fractals in the limit of $J\rightarrow +\infty$.

We have also studied the $F_{xy}$ model defined above, obtaining a
similar multifractal butterfly spectrum, as shown in  Fig.
\ref{Dqxy}(a). Interestingly, despite that the outline of the
butterfly spectrum of $F_{xy}$ is much similar to that for $F$,
careful investigations reveal considerable differences between the
butterfly spectrum of $F_{xy}$ and that for $F$.  For example, the
fractal dimension $D_q$ shown in Fig. \ref{Dqxy}(b) for $F_{xy}$ is
similar to, but slightly larger than, that of $F$ for $q>0$. It is
found that this is because the gaps in the butterfly spectrum of
$F_{xy}$ is more densely filled than that of $F$ (compare Fig.
\ref{Dqxy} (a) and Fig. \ref{sptm1}(b)).  In the next subsection we
will point out an even more fundamental difference between these two
systems.

\subsection{Level crossings}

In this subsection we study the level crossings in the butterfly
spectrum as the parameter $\hbar_{\eta}$ varies. Note first that
due to the above-mentioned symmetry
$\langle m|F|n\rangle=\langle -m|F|-n\rangle$,
the eigenstates of $F$ can be classified into $J$ eigenstates of
odd-parity and $J+1$ states of even parity. As such, we should
investigate the crossings between different-parity states and
between same-parity states.  In either case, computationally it is
found that the minimal distance in $\hbar_{\eta}$ between two level
crossings decreases sharply  with $J$. So even for a rather small
$J\sim 10$ it is already numerically demanding to identify all the
level crossings.

\begin{figure}
\vspace{-0.2cm}\epsfig{file=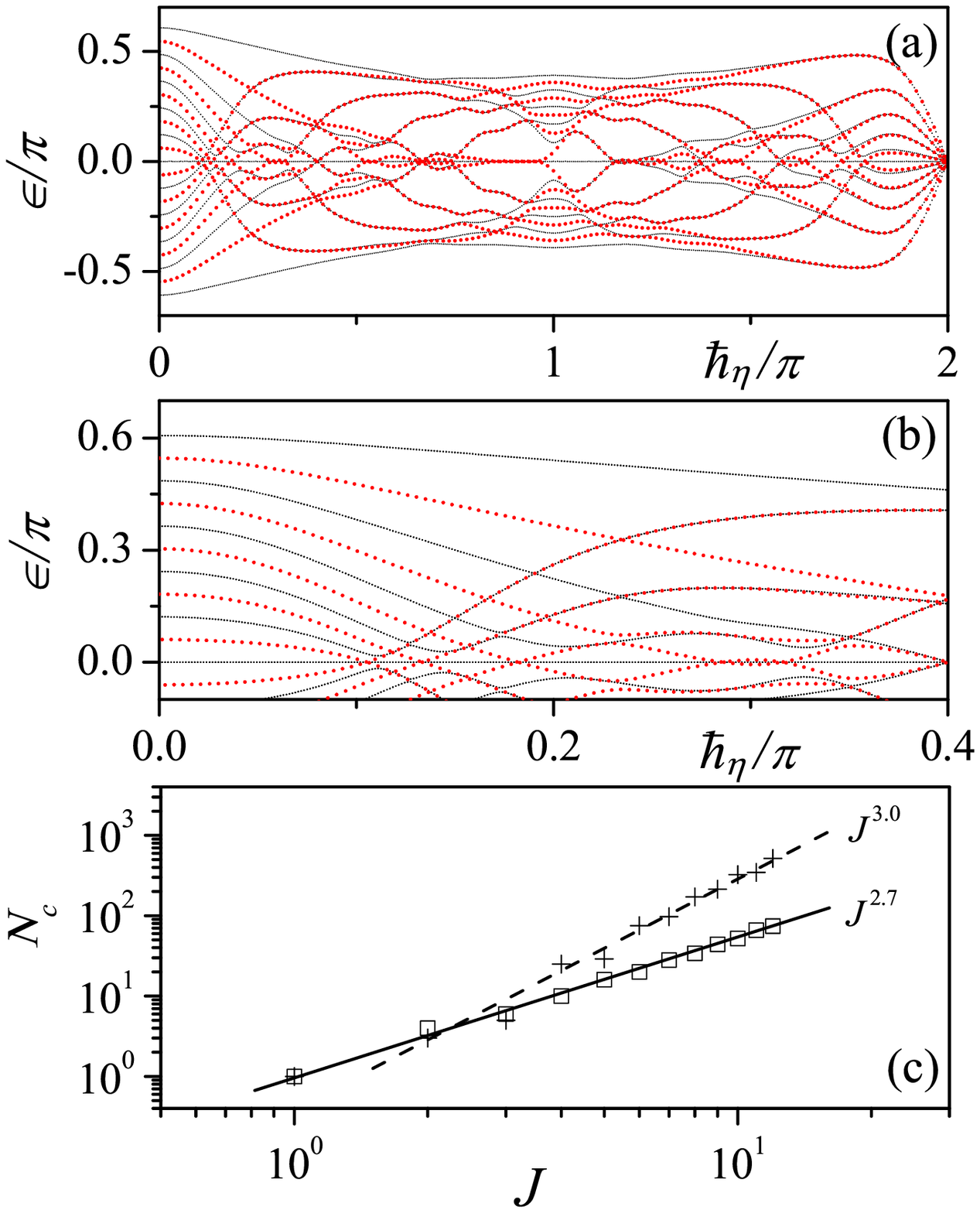,width=7.5cm}\vspace{-1.7cm}
\caption{(color online) (a) Level crossings between 10 even-parity
states (solid) and 9 odd-parity states (dotted) for $J=10$ and
$\alpha/\hbar_{\text{eff}}=1.0$. (b) A magnification of one part of
(a). (c) The number of level crossings versus $J$, for
$\hbar_\eta\in [0,4\pi)$ and $\alpha/\hbar_{\text{eff}}=1.0$.  The
cross (square) symbols are for crossings between different-parity
(same-parity) states and the fitting suggests a power law scaling
$J^{3.0}$ ($J^{2.7}$).} \label{crossing}
\end{figure}

As an example Fig. \ref{crossing}(a)-(b) presents the typical level
crossing behavior for $J=10$.  The Floquet states are seen to cross
each other frequently, between different-parity states and between
same-parity states. Both types of level-crossings turn out to be of
vast interest. For the first type, at a crossing point an arbitrary
superposition of two crossing states of different parities remains
an eigenstate but generally breaks the parity symmetry. So if such a
superposition state is used as the initial state, the ensuing
dynamics will maintain a nonzero population difference between the
two modes forever \cite{Korsch}. This makes it clear that the first
type of level crossings give rise to the seminal CDT phenomenon
\cite{Hanggi} that has attracted broad experimental and theoretical
interests.  It should be pointed out that in some regimes of
$\hbar_{\eta}$, to the naked eye two curves of opposite parities in
Fig. \ref{crossing}(a) and Fig. \ref{crossing}(b) are almost on top
of each other. As a result many CDT points are found in these
regimes. Note also that the CDT-induced population trapping is
fundamentally different from the well-known self-trapping effect on
the mean-field level. Indeed, the CDT effect here depends on $\eta$
and $J$, whereas mean-field self-trapping is transient and
independent of $J$.

\begin{figure}
\vspace{0.cm}\hspace{0.15cm}\epsfig{file=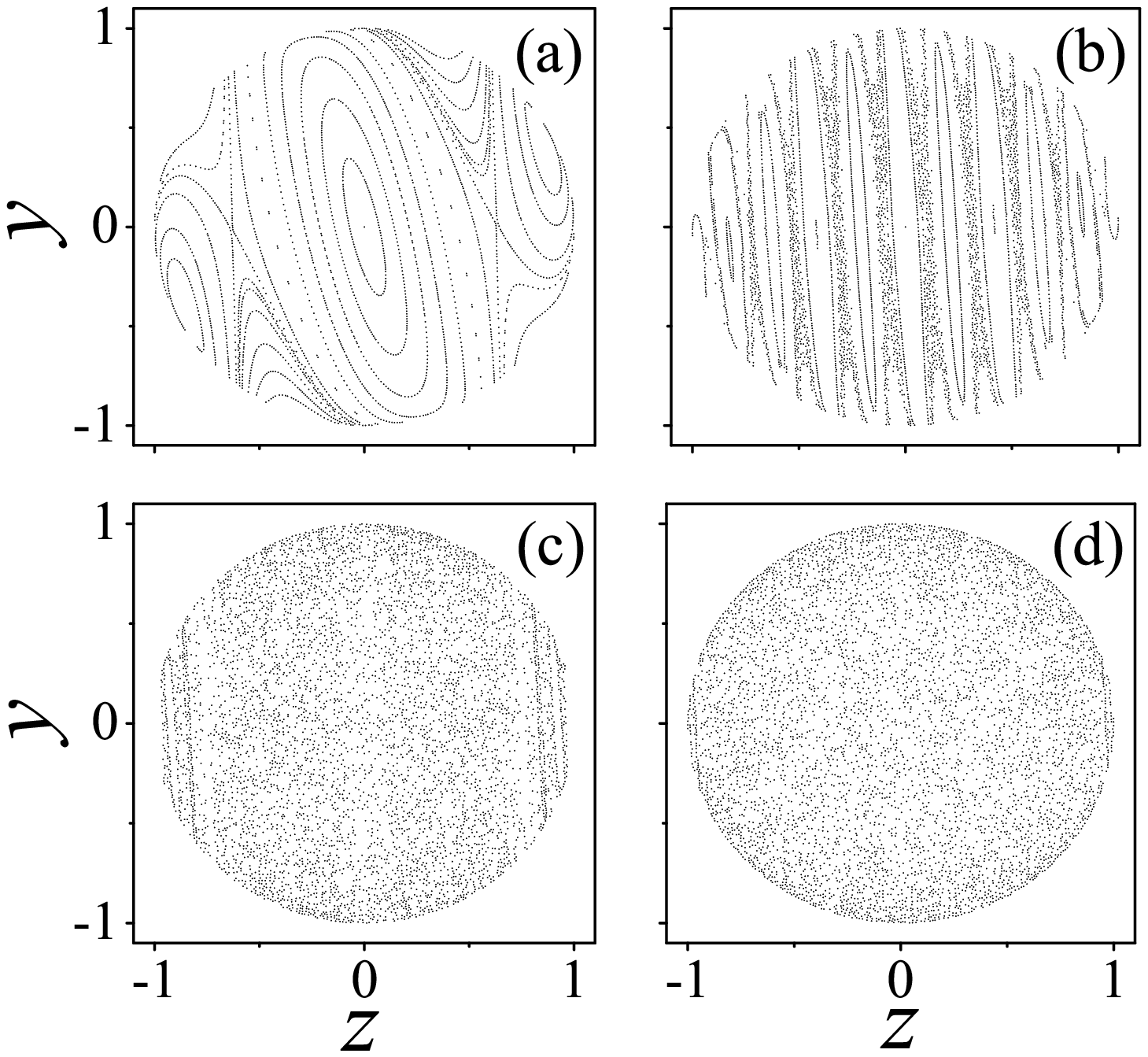,width=9.5cm}
\caption{Poincar\'{e} surfaces of section (with $J_x>0$) of the
classical or mean-field limit of $F$ in Eq. (3), with $\alpha=0.05$
(same as in Fig. 1(a)), $\eta=5$ in (a), 30 in (b), 75 in (c) and
100 in (d).}\label{class1}
\end{figure}

Now turning to the second type of level crossings, they come as a
surprise because avoided crossings
 between same-parity states, rather than true level crossings, are
generally anticipated for classically non-integrable systems (see
Figs. \ref{class1}-\ref{class2}). The second type of crossings
therefore suggest the uniqueness (e.g., some effective local
``symmetry") of $F$ whose matrix elements in the $|m\rangle$
representation are quasi-periodic. Recalling the above-mentioned
extreme example where all levels cross at $\hbar_\eta=2\pi$ for
integer $J$, we expect that special arithmetic properties of
$\hbar_{\eta}$ play a key role in both types of level crossings.

Careful checks are made to ensure that the same-parity level
crossings observed here are not avoided crossings with a very small
gap.  For example, we examined the crossing behavior for small $J$,
where analytical studies become possible. In particular, for $J=2$,
using Wigner's rotational matrices to express the 2nd and 4th
factors of $F$, we can analytically prove that there must be true
level crossings between two odd-parity states, at
$\hbar_{\eta}=2\pi/3$ and $\hbar_{\eta}=10\pi/3$, regardless of the
value of $\alpha$. This is fully consistent with our numerical
finding. Details for this case are presented in Appendix C. This
further confirms that the number theory properties of $\hbar_{\eta}$
are responsible for the same-parity level crossings. As another
check, we also studied the level crossing behavior in the butterfly
spectrum of $F_{xy}$. Therein we only obtain avoided crossings
between all the eigen-phases (see also Fig. 9 in Appendix C). For
$J\le 12$ and for the same parameters as in Fig. \ref{crossing}(c),
the typical gaps of the avoided level crossings in the spectrum of
$F_{xy}$ are found to be $> 10^{-6}$, many orders of magnitude
larger than the accuracy of our eigen-phase calculations
($10^{-13}$). This ``control case" hence indirectly supports our
observation of same-parity level crossings for $F$.

By obtaining all the level crossings in the butterfly spectrum of
$F$ with high accuracy for $J\leq 12$, we obtain in Fig.
\ref{crossing}(c) that the number of CDT points contained in the
butterfly patterns scales as $J^{3.0}$ and the number of same-parity
crossings scales as $J^{2.7}$. In either case, the number of
crossings divided by the total number of levels ($\sim J$) or
divided by the total number of level pairs ($\sim J^2$) diverges as
$J\rightarrow +\infty$. In particular, we assert that as $N$ goes to
infinity, on average each pair of Floquet states in a butterfly
pattern see infinite CDT points.

\subsection{Relevance of the classical limit}

An interesting question is what implications the classical dynamics
may have on the fractal spectrum observed in the $F$ system. To this
end we investigate the classical limit of $F$ defined above.
Numerical studies indicate that classically the system can be
governed by both regular and chaotic motions, depending on the two
system parameters $\eta$ and $\alpha$. In general, for a fixed
$\alpha$, as $\eta$ increases the classical dynamics undergoes a
transition from being regular to being chaotic. As an example Fig.
\ref{class1} presents the phase space structure of the classical
limit of $F$ with $\alpha=0.05$ and an increasing $\eta$.

Such a classical regular-to-chaos transition lacks a quantum
counterpart in the butterfly spectrum shown in Fig. \ref{sptm1} and
Fig. \ref{sptm2}, whose characteristics can be much similar for
radically different values of $\eta$. Indeed, the quantum Floquet
spectrum is periodic in $\eta$ with a period $4J\pi$. (See Appendix
A.) Therefore, upon quantization the regular or chaotic nature of
the classical dynamics may not necessarily be reflected in the
spectrum and hence can be irrelevant to the quantum dynamics.

\begin{figure}
\vspace{0.cm}\hspace{0.15cm}\epsfig{file=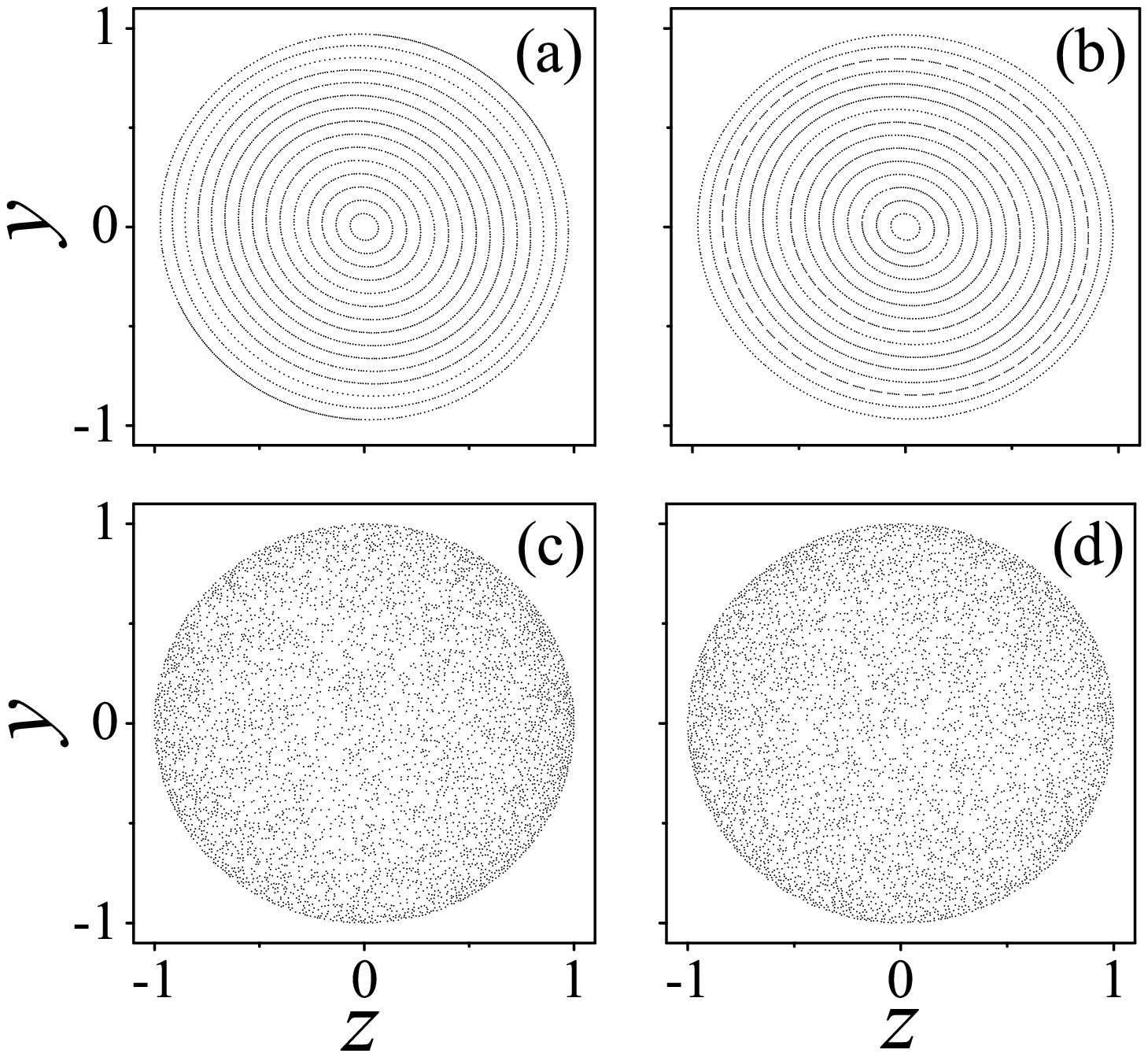,width=9.5cm}
\caption{Same as in Fig. \ref{class1} but $\alpha=0.1/3$ in (a) and
$\alpha=1/3$ in (b) with $\eta=0.06\pi$; $\alpha=0.1/3$ in (c) and
$\alpha=1/3$ in (d) with $\eta=0.06\pi+120\pi$. Quantum
mechanically, for $J=30$, cases (a) and (c), or cases (b) and (d),
share the same spectrum due to the spectral symmetry under
$\hbar_\eta \to \hbar_\eta+4\pi$. }\label{class2}
\end{figure}

Interestingly, the regular-to-chaos transition in the classical
dynamics is not even connected to the dissolving of the butterfly
spectrum (see Fig. \ref{sptm2}). As an example, let us examine such
a dissolving process for $J=30$ from $\alpha/\hbar_{\text {eff}}=1$
[Fig. \ref{sptm1}(b)] to $\alpha/\hbar_{\text {eff}}=3$, $6$ and
$10$ (Fig. \ref{sptm2}), corresponding to $\alpha=0.1/3, 0.1, 0.2$,
and $1/3$ respectively. Consider the spectrum along the vertical
cuts of these figures at $\hbar_{\eta}/\pi=2\times 10^{-3}$, i.e.,
$\eta=0.06\pi$. It is found that in spite of the drastic changes of
the spectrum in this dissolving process, the dynamics in the
classical limit remains regular all the way from $\alpha=0.1/3$ to
$1/3$ [see Fig. \ref{class2}(a)-(b)]. On the other hand, due to the
$4\pi$ periodicity of the quantum spectrum in $\hbar_{\eta}$, we
expect the same dissolving process if we change $\eta$ from
$0.06\pi$ to $0.06\pi+120\pi$. However, the corresponding classical
dynamics is now completely chaotic [see Fig. \ref{class2}(c)-(d)].
Hence the dissolving of the butterfly spectrum is unrelated to the
classical dynamics. This is in contrast to the kicked-Harper model,
where the dissolving of Hosftadter's butterfly was somewhat
connected to the classical regular-to-chaos transition
\cite{kickedharper}.

The little relevance of the classical limit in understanding the
butterfly spectrum also makes our model markedly different from the
conventional kicked-top model as a paradigm for quantum chaos. We
believe that such a lack of classical-quantum correspondence is a
result of a built-in feature of our SU(2) model. That is, our
strategy for generating the fractal spectrum is based on a partial
cancelation of quasi-random dynamical phases (see Eq.
\ref{rdmphase}), and this partial cancelation of quantum phases is a
pure quantum feature with no classical analog.  To double check this
we also studied the classical dynamics of the $F_{xy}$ model and
reached the same conclusion.

\subsection{Generating a fractal-spectrum family}

In the previous subsection we have studied the butterfly spectrum
associated with $F$. Here we point out that the butterfly spectrum
of $F$ is just one member of a whole butterfly-spectrum family. Let
us restrict ourselves to cases of integer $J$ and consider the
following function
\begin{eqnarray}
g(t)=\sum_{n}[g_0\delta(t-2n\tau-\tau)+\tilde
g_0\delta(t-2n\tau)]\nonumber
\end{eqnarray}
instead of that used in generating $F$ as in Eq. (\ref{Feq}).  Here
$\tilde g_0=(\frac{\pi}{2}\cdot\frac{\nu}{\mu}-g_0)$ and $\mu$ and
$\nu$ are two integers sharing no common factors. We can then obtain
an extended class of Floquet operators
\begin{eqnarray}
F^{(\nu/\mu)}\equiv  e^{i2\pi J_z^2 \nu/\mu} e^{i\eta
J_z^2/(2J)}e^{-i
\alpha J_x} e^{-i\eta J_z^2/(2J)} e^{-i\alpha J_x}\nonumber\\
= e^{i2\pi J_z^2
\nu/\mu}F.~~~~~~~~~~~~~~~~~~~~~~~~~~~~~~~~~~~~~~~~\label{Fnumu}
\end{eqnarray}
We also note that $F^{(\nu/\mu)}$ can be obtained if we set
$4g_0\tau/\xi=2\pi\nu/\mu$ in Eq. (\ref{2ndF}).  Obviously, $F$
corresponds to the special case of $\nu=\mu=1$.

In the double-kicked-rotor model considered in Ref. \cite{jiao09},
one can construct analogous operators by employing high-order
quantum resonances, where $\nu/\mu$ indicates the resonance order.
For each choice of $\nu/\mu$ a certain type of fractal spectrum can
be generated. An example for $\nu/\mu=1/2$, the so-called
anti-resonance condition, can be found in Ref. \cite{jiao09} (see
Fig. 3 therein).

Interestingly,  the spectrum of $F^{(\nu/\mu)}$ defined in Eq.
(\ref{Fnumu}) also forms a fractal-spectrum family, with the outline
of each butterfly roughly similar to its relative in the
double-kicked-rotor model considered in Ref. \cite{jiao09}.  We also
find that for integer $J$, the peculiar spectral characteristics of
$F$ discussed above can be maintained in the spectrum of
$F^{(\nu/\mu)}$. For example, we have investigated thoroughly the
case of $\nu/\mu=1/2$ and obtained that (1) $D_q$ curve is
qualitatively the same as that for $F$; i.e. $D_0=1$ and $D_q$
decreases as $q$ increases, (2) level crossings between eigenstates
of the same or different parities can occur; and (3) the dynamics in
the classical limit is also irrelevant in understanding the quantum
spectrum. Similarly, the spectral properties of $F_{xy}$ are also
found to be similar to its high-order extension
$F^{(\nu/\mu)}_{xy}\equiv e^{i2\pi J_z^2 \nu/\mu}F_{xy}$.  

The possibility of constructing such a butterfly-spectrum family
provides further support that our strategy for generating a fractal
Floquet spectrum in driven quantum systems is quite general.

\section{Discussion}

Experimental confirmation of a butterfly Floquet spectrum in driven
systems is challenging. In the case of the kicked-Harper model,
there have been a few experimental proposals but so far the
kicked-Harper model has not been experimentally realized.  The
double-kicked-rotor model proposed in Ref. \cite{jiao08} opens up a
new opportunity. However, in atom-optics realizations of the
kicked-rotor model, a dilute cold gas in a kicking one-dimensional
optical-lattice potential is required,  the quasi-momentum spread of
the initial state should be sufficiently narrow, and the interaction
between the atoms should be negligible. By contrast, in the present
study, we rely on an interacting cold gas distributed on two modes,
and there is no quasi-momentum issue. Moreover, because the
effective Planck constant $\hbar_\eta$ is simply given by $8g_0$, by
tuning the atom-atom interaction constant alone we may scan the
butterfly spectrum already. These advantages make the systems
proposed in this study one more step closer to possible experiments
of a butterfly Floquet spectrum.

\begin{figure*}
\vspace{0.cm}\hspace{0.cm}\epsfig{file=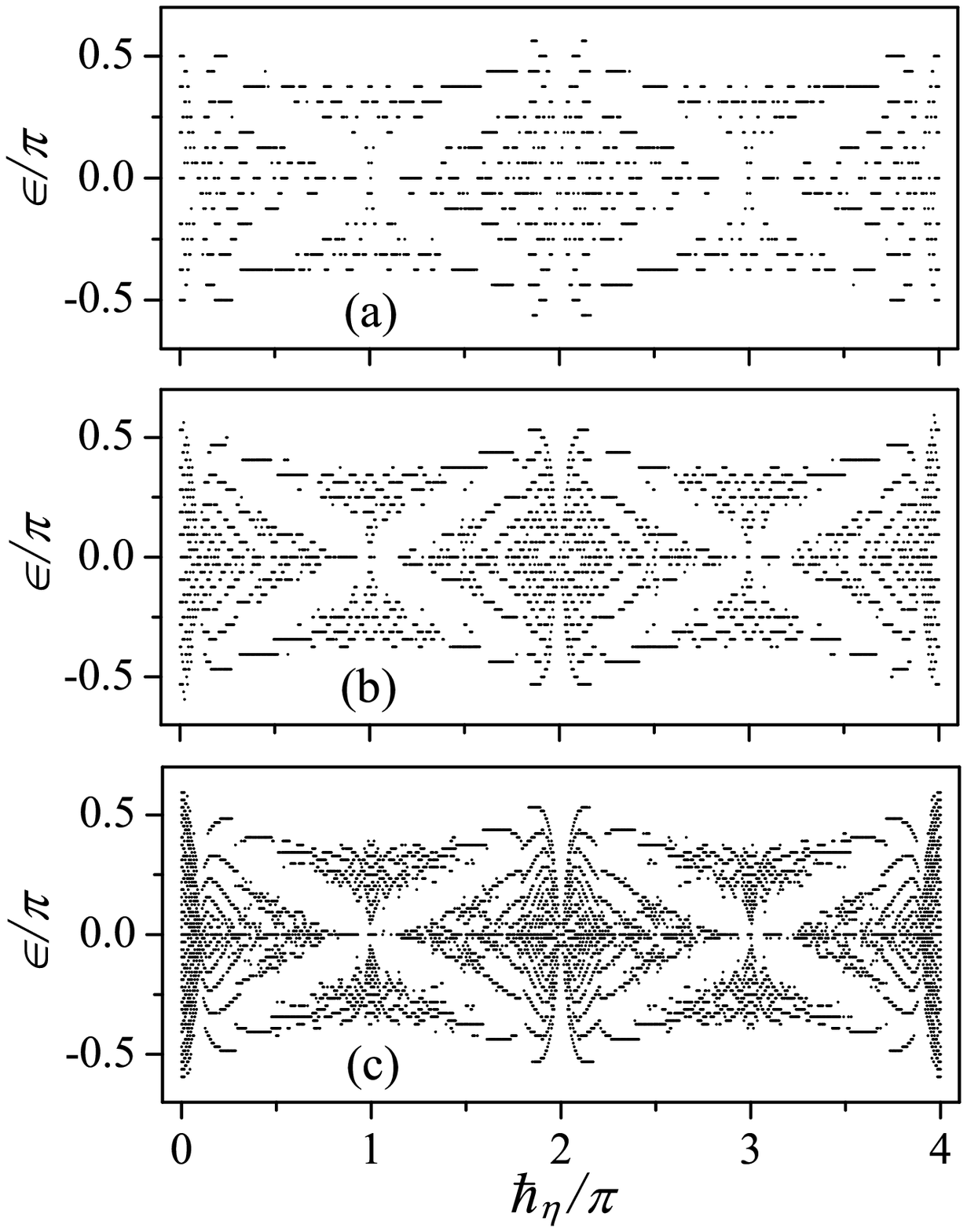,width=7.5cm}
\vspace{0.cm}\hspace{0.cm}\epsfig{file=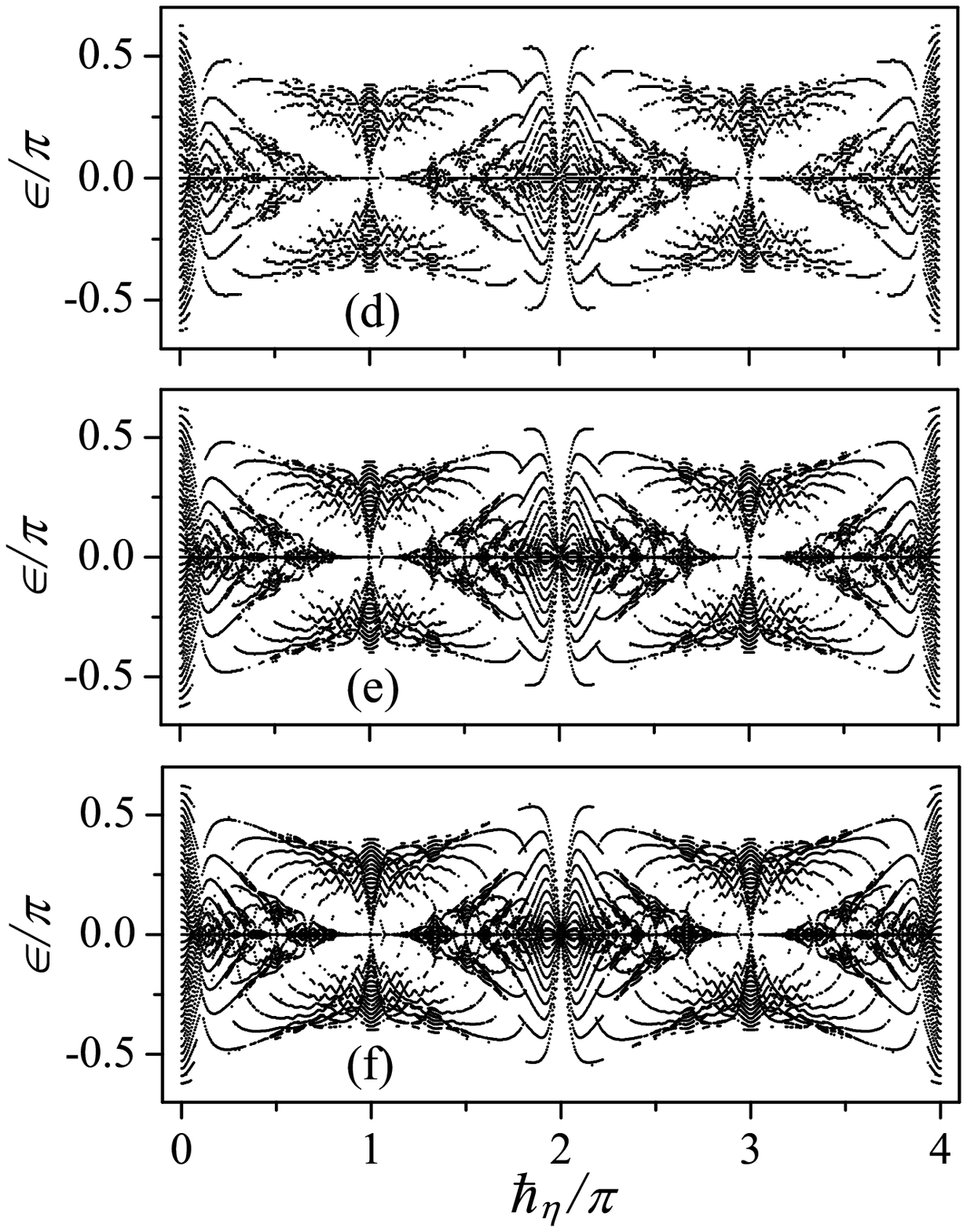,width=7.5cm}
\vspace{-1.5cm}\caption{Eigen-phase spectrum of the Floquet operator
$F$ for $J=20$ and $\alpha/\hbar_{\text{eff}}=1$,  retrieved via an
FFT of the sequence $\{\langle \phi(0)|F^n|\phi(0)\rangle,
n=1,\cdots,N\}$ with the sequence length $n=32 (a), 64 (b), 128(c),
256(d), 512(e)$ and $1024(f)$. The initial condition is given by
$|\phi(0)\rangle=|10\rangle$ as an example.}\label{FFT}
\end{figure*}

There are also other motivations for further theoretical and
experimental studies of our model. First, because the found
butterfly spectrum collapses at $\hbar_{\eta}=2\pi$ (or $g_0=\pi/4$)
for integer $J$, one may experimentally determine if $N$ is even or
odd by scanning the dynamics in the neighborhood of $g_0=\pi/4$.
Note that detecting the even-odd properties of $N$  is impossible in
the mean-field dynamics of a BEC, and such a topic is already under
investigation in Ref. \cite{oddeven} using a different mechanism.
Similarly, one may study the CDT points to reveal non-mean-field
effects. Second, it is now of great interest, both experimentally
and computationally, to revisit early results of how a multifractal
spectrum can be manifested in time-dependent properties
\cite{dynamics}. Third, noticing that recently dissipative two-mode
BEC's have attracted considerable attention \cite{dissp}, it seems
interesting to study dissipation effects on a butterfly spectrum.

Experiments of our model system with a small $J$ can be also
interesting. For example, for the $J=2$ case we have found that the
$F$ operator is an identity matrix in the odd-parity subspace at
$\hbar_\eta=2\pi/3$ and $10\pi/3$. (See Appendix C).  Such cases may
be used to extract the value of $g_0$ to a good precision. They may
be also useful to explore the implication of level crossings between
same-parity states.

We next discuss how to experimentally confirm a butterfly spectrum.
Certainly, a Floquet spectrum cannot be directly measured. However,
in principle there is a standard procedure to invert the Floquet
spectrum from the dynamics. This involves quantum state
reconstruction of the evolving two-mode BEC. That is, after each
period of the driving field, enough measurements on different
observables are to be made to reconstruct the state of the evolving
two-mode BEC.  This is possible because quantum state reconstruction
of a BEC is currently a very active and fruitful area. This state
reconstruction is not expected to be prohibitively demanding if $N$
is relatively small. With the quantum states at different times
reconstructed, then  by making a Fast-Fourier-Transformation (FFT)
of the time-evolving state, the Floquet spectrum may be obtained.

We point out that there is no need to follow the dynamics for very
long in order to resolve a butterfly spectrum. Let us assume that
the wavefunctions are already reconstructed by sufficient
measurements at integer multiples of the driving period $\tau$. To
examine how many periods are needed for obtaining the butterfly
spectrum from experiments,  we consider an example where the initial
state is chosen as $|m=10\rangle$ for $J=20$. Using the same
parameters as in Fig. \ref{sptm1}(a), and using the time-evolving
states after $n=32$, $64$, $128$, $256$, $512$ and $1024$ kick
pairs, the spectrum obtained via FFT are shown in Fig. \ref{FFT}. It
is seen that a few hundred kicks can be good enough to resolve the
shape of a butterfly spectrum relatively well. If, as discussed in
the following, the time scale of the kicking period is chosen to be
$10^{-5} s$, then this means that the required duration to follow
the dynamics in an experiment is around $0.01 s$.

Finally, let us comment on the parameter ranges we have chosen.
First of all, to ensure that a wide regime of the found butterfly
spectrum can be visited in a real system, a tunable $g_0$ is
required, and its characteristic value should be $\sim 1$. Let $g_c$
be the self-interaction constant of a static two-mode BEC. A
reasonable range of $g_c/f$ ($f$ is the tunneling rate) is from
$10^{-3}$ to $10^{-2}$ for a two-mode BEC in a double-well potential
\cite{Milburn}. With the first realization of $F$ in mind and for
$\alpha J=1$ and  $J=10$, we have $g_c/f=2\tau g_c/\alpha$, and that
$2\tau g_c$ ranges from $10^{-4}$ to $10^{-3}$.  This indicates that
$g_0/(\tau g_c)$ should be around $10^{3}-10^{4}$. If the value of
$g_c$ (in SI unit of frequency) is about $50\ s^{-1}$ (a value
considered in \cite{Milburn}), we have that $\tau$ is in the range
of $10^{-6} s$ to $10^{-5} s$. Considering other realizations of a
two-mode BEC might lead to different characteristic values of $\tau$
and $g_0/(\tau g_c)$.

\section{Conclusion}

We have presented a strategy for generating a fractal butterfly
Floquet spectrum in a class of driven SU(2) systems. The essence of
this strategy is to partially cancel the quasi-random dynamical
phases in the time evolution and then induce intriguing spectral
properties. The success of such a strategy in both this work and our
early work \cite{jiao08} treating a double-kicked-rotor model
indicates its wide applicability. As such, butterfly Floquet
spectrum is expected to occur in many driven quantum systems that
can go well beyond the context of two-dimensional electron systems
or the Harper model context.  A butterfly spectrum thus obtained may
also differ significantly from Hofstadter's butterfly.


Detailed aspects of the butterfly spectrum in driven SU(2) systems
are also examined. The level crossing features in the butterfly
spectrum, especially its connection with coherent destruction of
tunneling, and the surprising observation of level crossings between
same-parity states, are emphasized.  The sensitivity of the
butterfly spectrum on the even-odd properties of the number of
particles is also highlighted.  Many further questions can be asked
and we believe that our driven SU(2) model will be relevant to a
number of related research topics, including fractal statistics,
quantum chaos, quantum control, and the validity of mean-field
descriptions of Bose-Einstein condensates.

The conventional system for understanding a butterfly Floquet
spectrum is the kicked-Harper model. Therein the quantization rule
varies with the boundary condition adopted \cite{guarneri} and a
compact toroidal phase space arises only if the Planck constant
assumes special values \cite{Leboeuf}. A general treatment of the
kicked-Harper model leads to a band structure that often
complicates the issue. By contrast, the phase space structure in
our driven SU(2) model is necessarily on a sphere
\cite{Haakebook}, with no arbitrariness in quantization and no
band structure in the spectrum. For these reasons we hope that the
butterfly Floquet spectrum discovered here can stimulate future
studies on general implications of a fractal Floquet spectrum.
Combining this work with our early study \cite{jiao08}, we have
that three paradigms of quantum and classical chaos, i.e., the
kicked-rotor model, the kicked-Harper model, and the kicked-top
model, are linked all together because any one of the three
contexts can be used to generate quantum critical systems with a
fractal Floquet spectrum.






\section*{Acknowledgements}

We thank Prof. C.-H. Lai for his kind support and for making this
collaborative work possible.  J.W. acknowledges support from DSTA of
Singapore under agreement of POD0613356. J.G. is supported by WBS
grant Nos. R-144-050-193-101/133 and the NUS ``YIA" (grant No.
R-144-000-195-101).

\appendix

\section{$4\pi$ periodicity of $F$ in $\hbar_\eta$}

Denoting by $U(\hbar_\eta)\equiv e^{\eta J_z^2/(2J)}$, the
operator $F$ can be written as $F(\hbar_\eta)= U(\hbar_\eta)e^{-i
\alpha J_x} U^\dag (\hbar_\eta) e^{-i\alpha J_x}$. It can be shown
that
\begin{eqnarray}
U(\hbar_\eta+4\pi)=\pm U(\hbar_\eta),\label{U}
\end{eqnarray}
hence $F(\hbar_\eta+4\pi)=(\hbar_\eta)$. To show the above equality
let us consider the representation using the eigenstates of $J_z$,
i.e. $J_z|m\rangle=m\hbar|m\rangle$, in which $U$ is diagonalized
and $\langle
m|U(\hbar_\eta+4\pi)|m\rangle=e^{i(\eta+4J\pi)m^2/(2J)}$, resulting
in $U(\hbar_\eta+4\pi)= U(\hbar_\eta)$ if $J$ is an integer, and
$U(\hbar_\eta+4\pi)= -U(\hbar_\eta)$ if $J$ is a half-integer. In
the latter case as $U^\dag(\hbar_\eta+4\pi)= -U^\dag(\hbar_\eta)$,
which contributes a second minus sign, we again have
$F(\hbar_\eta+4\pi)=F(\hbar_\eta)$.

\section{Spectrum symmetry of $F$ under $\hbar_\eta\to -\hbar_\eta$}

Consider first the $F$ operator and its eigenfunction
$|\psi\rangle$ with
\begin{eqnarray}
F(\hbar_\eta) |\psi\rangle = e^{-i\omega} |\psi\rangle.\label{A1}
\end{eqnarray}
Multiplying $e^{-i\hbar_\eta J_z^2/2}e^{-i\alpha J_x}$ to both
sides of the above eigenfunction equation and defining that
$|\psi'\rangle\equiv e^{-i\hbar_\eta J_z^2/2} e^{-i\alpha
J_x}|\psi\rangle$, then we have
\begin{eqnarray}
F(-\hbar_\eta) |\psi'\rangle = e^{-i\omega}
|\psi'\rangle.\label{A2}
\end{eqnarray}
Hence, comparing Eqs. (\ref{A1}) and (\ref{A2}) one sees that the
spectrum of $F(\hbar_\eta)$ is identical with that of
$F(-\hbar_\eta)$.

Combining this result with the $4\pi$ periodicity of $F$ in
$\hbar_\eta$, one obtains that the butterfly spectrum of $F$ has a
reflection symmetry with respect to $\hbar_\eta=2\pi$.

\section{Level crossings between two odd-parity states of $F$
for $J=2$}

The arithmetic properties of $\hbar_\eta$ can cause surprising and
frequent level crossings between same-parity eigenstates  of $F$. To
understand this counter-intuitive result we consider a simple case
with $J=2$. Numerically we observe level crossings between two
odd-parity states at $\hbar_\eta=2\pi/3$ and $\hbar_\eta=10\pi/3$
besides that at $\hbar_\eta=2\pi$. (see Fig. 9a for
$\alpha/\hbar_{\text{eff}}=1$). Here we give a mathematical proof
that this is indeed the case, thus supporting the numerically
observed level crossings in general. Note that our approach can be
extended to other cases as well.

\begin{figure}
\vspace{0.cm}\hspace{0.cm}\epsfig{file=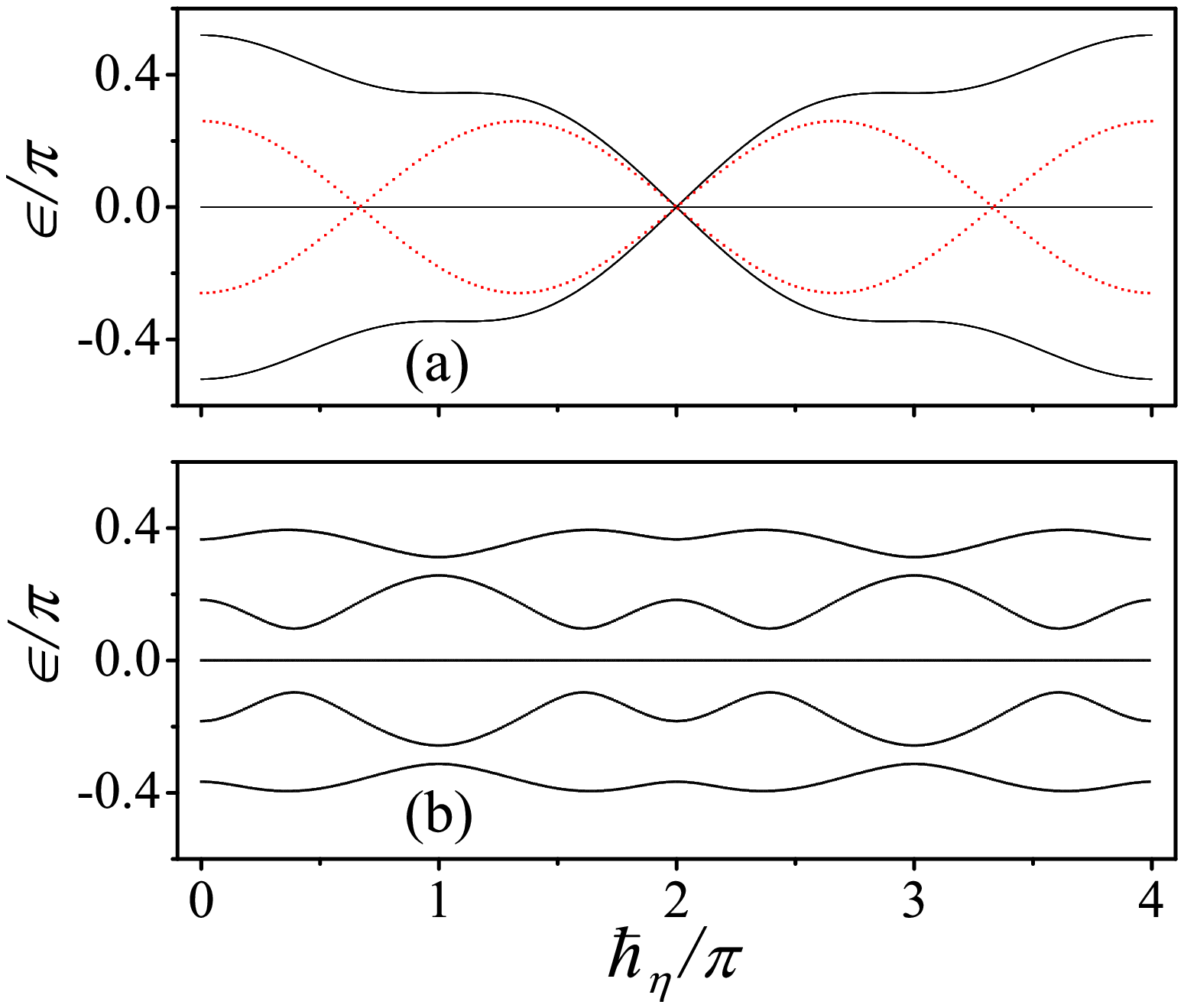,width=9.5cm}
\vspace{-.5cm}\caption{(color online) The eigen-phase spectrum of
the Floquet operator $F$ (a) and $F_{xy}$ (b) for $J=2$ and
$\alpha/\hbar_{\text{eff}}=1$. In (a) the two eigen-phases of odd
parity (dotted) cross at $\hbar_{\eta}=2\pi/3, 2\pi$, and $10\pi/3$,
while the three eigen-phases of even parity (solid) only cross at
$\hbar_{\eta}=2\pi$. As a contrast, the spectrum of $F_{xy}$
displays avoided crossings instead.}\label{figA1}
\end{figure}

We first note that each of the four factors of $F$ preserves the
parity. So if we consider the odd-parity subspace only, then each
of the four factors can be reduced to a $2\times 2 $ matrix
because the odd-parity subspace is two-dimensional for $J=2$.
Specifically, we consider the following basis states
\begin{eqnarray}
|\tilde{2}\rangle&\equiv &(|2\rangle-|-2\rangle)/\sqrt{2} \nonumber \\
|\tilde{1}\rangle&\equiv &(|1\rangle-|-1\rangle)/\sqrt{2}
\nonumber
\end{eqnarray}
In this representation, the factor $e^{i\hbar_\eta J_z^2/2}$ and
$e^{-i\hbar_\eta J_z^2/2}$ for $\hbar_\eta=2\pi/3 $ are given by
\begin{eqnarray}
U\equiv\left(\begin{array}{cc} e^{i4\pi/3} & 0 \\ 0 &  e^{i\pi/3} \\
\end{array}\right) \nonumber
\end{eqnarray}
and $U^\dag$ respectively.

In order to get the matrix expression of $F$, we need to get the
analytical expression for the factor $e^{-i\alpha J_x}$.
Fortunately, this can also be done by using Wigner's rotational
matrices \cite{Zare}. Denoting $D^{2}_{m'm}$ as Wigner's expression
for the rotational matrix, we have
\begin{eqnarray}
D^{2}_{m'm}=\langle m'| e^{-i\alpha J_y}| m\rangle.
\end{eqnarray}
What we need is the matrix $\langle m'| e^{-i\alpha J_x}|
m\rangle$, which is related to $D^{2}_{m'm}$ by
\begin{eqnarray}
\langle m'| e^{-i\alpha J_x}| m\rangle=i^{m'-m}D^{2}_{m'm}.
\end{eqnarray}
Using the explicit expressions of the rotational matrix
$D^{2}_{m'm}$ given in \cite{Zare}, we obtain
\begin{eqnarray}
\langle 2| e^{-i\alpha J_x}| 2\rangle &=& \cos^4(\alpha/2)\nonumber \\
\langle 2| e^{-i\alpha J_x}| -2\rangle &=& \sin^4(\alpha/2) \nonumber \\
\langle 1| e^{-i\alpha J_x}| 1\rangle &=&
[1+\cos(\alpha)][2\cos(\alpha)-1]/2 \nonumber \\
\langle 1| e^{-i\alpha J_x}| -1\rangle &=&
-[1+2\cos(\alpha)][1-\cos(\alpha)]/2 \nonumber \\
\langle 2| e^{-i\alpha J_x}| 1\rangle &=& -i
\sin(\alpha)[1+\cos(\alpha)]/2 \nonumber \\
\langle 2| e^{-i\alpha J_x}| -1\rangle &=& -i
\sin(\alpha)(\cos(\alpha)-1)/2 \nonumber \\
\langle 1| e^{-i\alpha J_x}| 2\rangle &=& -i
\sin(\alpha)[\cos(\alpha)+1)]/2 \nonumber \\
\langle 1| e^{-i\alpha J_x}| -2\rangle &=& -i
\sin(\alpha)[\cos(\alpha-1)]/2. \nonumber
\end{eqnarray}
In the odd-parity subspace,  we then find
\begin{eqnarray}
\langle\tilde{2}|e^{-i\alpha J_x}|\tilde{2}\rangle &=&
\cos(\alpha)
\nonumber \nonumber \\
\langle\tilde{2}|e^{-i\alpha J_x}|\tilde{1}\rangle &=&
-i\sin(\alpha) \nonumber\\
\langle\tilde{1}|e^{-i\alpha J_x}|\tilde{2}\rangle &=&
-i\sin(\alpha)
\nonumber \\
\langle\tilde{1}|e^{-i\alpha J_x}|\tilde{1}\rangle &=&
\cos(\alpha). \nonumber
\end{eqnarray}
Finally, the $F$-matrix in the odd-parity subspace is given by
\begin{eqnarray}
F&=&U e^{-i\alpha J_x} U^\dag e^{-i\alpha J_x}
=\left(\begin{array}{cc} 1 & 0 \\ 0 &  1 \\
\end{array}\right).\nonumber
\end{eqnarray}
Because $F$ turns out to be a unit matrix,  this directly
demonstrates that the two eigenphases of $F$ are both zero
(because $e^{i0}=1$).  Two odd-parity eigenstates hence cross each
other at $\hbar_\eta=2\pi/3$.  The symmetry of the spectrum then
indicates another crossing at $\hbar_\eta=10\pi/3$. In addition,
this proof shows that this crossing is independent of $\alpha$,
which is also confirmed by our numerical calculations.  Our proof
can also be extended to other cases with a rather small $J$,
because thanks to Wigner, the rotational matrix elements can be
analytically obtained.

\end{document}